\documentclass[draftcls,12pt,onecolumn]{IEEEtran}
\usepackage{epsfig}
\usepackage{amsmath,amssymb}
\usepackage{moreverb}
\usepackage[colorlinks,bookmarksopen,bookmarksnumbered,citecolor=red,urlcolor=red]{hyperref}

\newcommand\BibTeX{{\rmfamily B\kern-.05em \textsc{i\kern-.025em b}\kern-.08em
T\kern-.1667em\lower.7ex\hbox{E}\kern-.125emX}}

\begin{document}

\title{Geometric projection-based switching policy for multiple energy harvesting transmitters
\thanks{This work was supported by the National Natural Science Foundation of China (61162008, 61172055), the Guangxi Natural Science Foundation (2013GXNSFGA019004), the Open Research Fund of State Key Laboratory of Networking and Switching Technology (SKLNST-2011-1-01), the Open Research Fund of Guangxi Key Lab of Wireless Wideband Communication \& Signal Processing (12103), the Director Fund of Key Laboratory of Cognitive Radio and Information Processing (Guilin
University of Electronic Technology), Ministry of Education, China (2013ZR02), and the Innovation Project of Guangxi Graduate Education (YCSZ2012066).}}

\author{Hongbin Chen, Fangfang Zhou, Jun Cai, Feng Zhao, and Qian He
\thanks{H. Chen, F. Zhou, F. Zhao, and Q. He are with the Key Laboratory of Cognitive Radio and Information Processing (Guilin University of Electronic Technology), Ministry of Education, Guilin, China.}
\thanks{H. Chen and Q. He are with the State Key Laboratory of Networking and Switching Technology (Beijing University of Posts and Telecommunications), Beijing, China.}
\thanks{J. Cai is with the Department of Electrical and Computer Engineering, University of Manitoba, Winnipeg, Manitoba, Canada.}
}

\maketitle

\begin{abstract}
Transmitter switching can provide resiliency and robustness to a communication system with multiple energy harvesting
transmitters. However, excessive transmitter switching will bring heavy control overhead. In this paper, a geometric
projection-based transmitter switching policy is proposed for a communication system with multiple energy harvesting
transmitters and one receiver, which can reduce the number of switches. The results show that the proposed transmitter
switching policy outperforms several heuristic ones.
\end{abstract}


\section{Introduction}
Along with the advancement of wireless communication technologies
and energy harvesting devices, energy harvesting communication
systems attract great research attention in recent years
\cite{epiw}--\cite{gw}. However, in an energy harvesting communication
system, the energy that can be harvested from the environment is
unstable and varies over time. Thus, in order to guarantee the
quality-of-service requirements, such harvested energy has to be
used carefully.

Transmission completion time minimization is an important goal for energy harvesting communication systems. In literature, many transmission scheduling schemes to achieve this goal or the dual goal of throughput maximization under a given deadline have been developed by considering different scenarios, such as point-to-point communication \cite{opsi1,ty}, broadcasting
\cite{bwae}--\cite{eou}, finite battery \cite{obsf}, fading channels
\cite{tweh}, multiple access channel \cite{yu}, parallel broadcast
channels \cite{otsf}, two-user Gaussian interference channel \cite{ty2}, time varying channels \cite{ke}, wireless energy transfer \cite{aacu}, Markovian energy harvesting \cite{bgd}, energy storage losses \cite{dg}, and packet arrivals during transmission \cite{ouu}. However, none of these works considered transmitter switching when there are multiple transmitters available.

Transmitter switching has been discussed in an opportunistic relaying scheme
\cite{xb}. With transmitter switching, resiliency and robustness of
a communication system can be improved, especially when a
transmitter fails or is not in the best condition. However,
excessive transmitter switching will bring heavy switching control
overhead and transmission interruption. Therefore, a
transmitter switching policy should be designed to reduce
the number of switches and switching control overhead.

In our earlier work \cite{tsfb}, a transmitter switching policy for a
broadcast communication system with energy harvesting
transmitters was proposed. It focused on the case of two
transmitters and ignored the number of switches.
In this letter, a geometric projection-based transmitter switching policy for a communication
system with multiple energy harvesting transmitters and one receiver is proposed,
towards the goal of reducing the number of
switches. First, regarding the transmitters as a whole \cite{tsfb}, given the amount of bits to be sent, the transmission completion time is obtained in a deterministic manner \cite{opsi1}. Then, the time-data plane is constructed with the transmission completion time and the amount of bits to be sent. To complete data transmission with less number of switches, transmitter switching should follow the straight line connecting the transmission start point and the transmission completion point. With this criterion at hand, geometric projection is
applied to find the suitable transmitter to work. To our knowledge, no transmitter switching policy has been proposed for energy harvesting communication systems. Therefore, the proposed transmitter switching policy is compared with several heuristic ones to show its merit.

\section{Energy Harvesting Communication System Model\label{sec2}}
\begin{figure}[t]
\centerline{\hbox{\epsfxsize=7cm\epsfbox{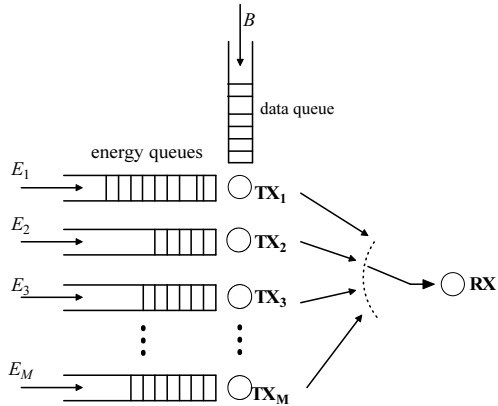}}}
\caption{Communication system with multiple energy harvesting
transmitters and one receiver.} \label{sm}
\end{figure}

We consider a communication system consisting of $M$ energy harvesting transmitters $\mbox{\rm TX}_1$, $\mbox{\rm TX}_2$, \ldots, $\mbox{\rm TX}_M$ and one receiver $\mbox{\rm RX}$, as shown in Figure~\ref{sm}.
Each transmitter has an energy queue while all the transmitters
share the same data queue (This kind of communication systems have wide potential applications. Take cellular networks for example: several base stations are deployed in a wild environment where electric power supply is unavailable. They harvest solar energy from the environment. If the data transmission for a user is not completed by a base station, the transmission can be handed over to another base station in order to provide uninterrupted data transmission service). A transmitter switching policy will be designed to choose one of the transmitters to send data at every switching moment. This can be done in a centralized or distributed fashion, which may bring some control overhead. The energies arriving
to the transmitters are stochastic and independent of each other.
It is assumed that the arriving time of every energy harvesting process
obeys Poisson distribution \cite{leht} and the amount of harvested energy in the arriving time of every energy harvesting process obeys uniform distribution (Since there is no model available in literature on the distribution of the amount of harvested energy, for explanation purpose, uniform distribution is considered in this paper. However, the proposed switching policy and the analysis procedure hold for other possible distributions). Moreover, the distribution of the amount of harvested energy does not depend on the distribution of the arriving time of every energy harvesting process. For the $m$-th transmitter $\mbox{\rm TX}_m$, the length of the energy
harvesting time slot $\mathcal{T}_{m,n}$ obeys
exponential distribution with parameter $\lambda_m$ and the amount
of harvested energy $E_{m,n}$ obeys uniform distribution in the
interval $(\mbox{\rm dn}_m, \mbox{\rm up}_m)$. Note that these parameters can be the same or different for different transmitters. It is assumed that
all data bits have arrived and are ready at the transmitters before
the transmission starts \cite{opsi1}. Energies harvested by the
transmitters are used for sending the data. It is assumed that the
batteries have sufficient capacity and the harvested energy will not
overflow. At every moment only one transmitter is
sending data. When this transmitter uses up its energy, another
transmitter will turn to send data. The switched transmitter may be
chosen to work again later. In the following, data transmission will
be analyzed from the information-theoretic point of view. It is
assumed that transmission is always successful no matter which
transmitter works. Moreover, the transmitters will not send
duplicated data.

Each transmitter sends data to $\mbox{\rm RX}$ through an additive
white Gaussian noise (AWGN) channel with path
loss. When $\mbox{\rm TX}_m$ works, the received signal can be
represented by \begin{equation}\label{eqyr1}
    y_{m}=h_m x+v_{m},\quad m=1,\,\cdots,\,M
\end{equation}
where $x$ is the transmitted signal, $h_m$ is the
path loss between $\mbox{\rm TX}_m$ and $\mbox{\rm RX}$, $v_{m}$ is
an AWGN with zero mean and variance $\sigma_m^2$. Without loss of
generality, we assume that
$\sigma_{1}^2=\sigma_{2}^2=\cdots= \sigma_{M}^2=N_0
B_w $, where $N_0$ is the noise power spectral density and $B_w$ is the
bandwidth.

Suppose that $\mbox{\rm TX}_m$ sends data with power $\tilde{P}_m$. Because there is only one transmitter sending data at every time, the channel shown in Figure~\ref{sm} is actually a point-to-point AWGN channel. Then, the capacity region is
\begin{equation}\label{rate}
  r_{m}\leq B_w\log_2\bigg(1+\frac{\tilde{P}_m h_{m}}{N_0 B_w}\bigg),\quad m=1,\,\cdots,\,M
\end{equation}

\begin{figure}[t]
\centerline{\hbox{\epsfxsize=7.0cm\epsfbox{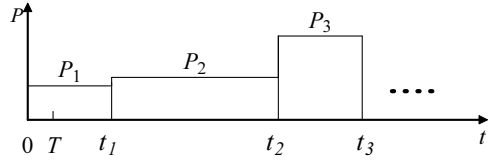}}}
\caption{Slot-by-slot optimal transmission power
of the whole transmitter.} \label{wh}
\end{figure}

Following our previous work in \cite{tsfb}, we treat the transmitters as a whole. The term ``whole transmitter'' means that we only care about the amount of bits that has been sent, rather than which transmitter does the transmission. Assuming the arriving time and the amount of harvested energy of every energy harvesting process are known and following the method in \cite{opsi1}, within the transmission completion time for a given amount of bits to be sent, we partition the time into slots and calculate the optimal transmission power in each time slot, as shown in Figure \ref{wh}. Note that each time slot may cover multiple energy harvesting moments and multiple transmitter switchings.
In Figure~\ref{wh}, $T$ is the present time, $t_j$, for $j=1,2,3\,\cdots$, is an
energy harvesting moment for a transmitter (called an epoch), and $P_j$ is
the optimal transmission power of the ``whole transmitter'' during
the time slot $[t_{j-1}, t_j]$ (Note that $P_j$ is adopted by any transmitter working
in this time slot). Following the method in \cite{opsi1}, at any
given time $T_e$, we can obtain the maximum amount of bits that can
be sent, denoted by $B_e$. In turn, if we set the amount of
bits to be sent as $B_e$, the time $T_e$ would be the
transmission completion time. When the ``whole transmitter'' works
with the optimal transmission power, the transmission completion
time reaches its minimum. After the minimum transmission completion
time is obtained, we design the transmitter switching policy which
intends to reduce the number of switches.

The problem of minimizing the number of switches can be formulated as follows:
\begin{align}
 \min_{l_{m,u}}\,\,&f=\sum_{m=1}^{M}f_m \nonumber\\
 s.t. \quad&\sum_{m=1}^{M}\sum_{u=1}^{f_m}l_{m,u}=T_e \nonumber \\
&l_{m,u}=\frac{E_{m,u}}{P_{l_{m,u}}},\enskip m=1,\cdots,M; u=1,\cdots,f_m\nonumber \\
&B_w\sum_{m=1}^{M}\sum_{u=1}^{f_m}l_{m,u}\log_2\bigg(1+\frac{P_{l_{m,u}} h_{m}}{N_0 B_w}\bigg)=B_e.
\end{align}
where $l_{m,u}$ is the length of the time slot for $\mbox{\rm TX}_m$ at the $u$th working with the available energy $E_{m,u}$ and working power $P_{l_{m,u}}$, $f$ is the total number of switches, and $f_m$ is the number of switches for $\mbox{\rm TX}_m$ before the transmission completion time $T_e$. Because the relationship between $f_m$ and ${l_{m,u}}$ cannot be obtained and $E_{m,u}$ depends on the length of the past working time slots, this optimization problem is very difficult to solve. Therefore, we resort to the geometric projection-based approach to reduce the number of switches.

\section{Geometric Projection-Based Transmitter Switching Policy\label{sec3} }
\begin{figure}[t]
\centerline{\hbox{\epsfxsize=5.9cm\epsfbox{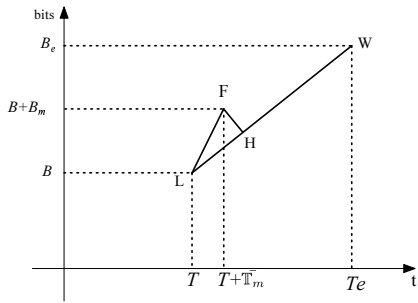}}}
\caption{Geometric projection on a straight line in the time-data
plane.} \label{tb}
\end{figure}

Our goal is to design a switching policy that can reduce the
number of switches within the transmission
completion time. The geometric projection approach is illustrated in
Figure~\ref{tb}.
For a given amount of bits to be sent $B_e$ and the
corresponding transmission completion time $T_e$, we plot them in a point
with label W. The point of the present time $T$ and the amount of
sent bits $B$ is plotted with label L. The point of the next switching moment $T+\bar{\mathbb{T}_{m}}$ and the amount of sent bits $B+B_m$
is plotted with label F, where $B_m$ is the amount of sent bits during the interval $[T, T+\bar{\mathbb{T}_{m}}]$.

As we know, a straight line is shortest between two points. Hence,
the fastest way to finish sending out $B_e$ bits is making the
transmission follow the line LW. To reduce the
number of switches, at the next switching moment
we will choose the transmitter corresponding to LH, which is the
longest projection of LF on LW (the projection is made for every transmitter except for the current working one). When the energy harvesting processes are unknown, to find out the transmitter that
corresponds to the longest projection, we have to predict the mean
of the working time of the transmitters (to predict LF) and the transmission completion point W in advance. In this case, it is impossible to use the real values of the
working time of the transmitters to perform projection. Therefore,
it is natural to use the average value instead. When the energy harvesting processes are known, the working time of the transmitters can be directly used and the transmission completion point can be obtained.

The proposed switching policy is summarized as
follows. First, the working time of each transmitter is calculated with the available energy and transmission power. Then, the amount of bits that can be sent within the working time for each transmitter is calculated and the points are plotted in the time-data plane. Finally, geometric projection on LW is done and the transmitter corresponding to the longest projection is chosen as the next working transmitter.

\begin{figure}[t]
\centerline{\hbox{\epsfxsize=8cm\epsfbox{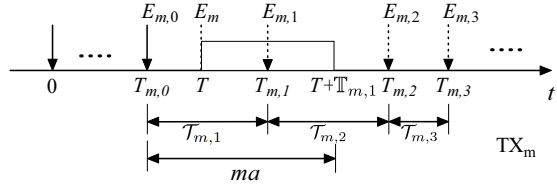}}} \caption{Illustration of predicting the mean of the working time for
$\mbox{\rm TX}_m$.} \label{tt}
\end{figure}

The process of predicting the mean of the working time of the
transmitters is described as follows (illustrated
in Figure~\ref{tt}). Assuming that in the interval
$[0,T]$ $\mbox{\rm TX}_1$ is working, we need to predict the mean
of the working time of $\mbox{\rm TX}_2$ to $\mbox{\rm TX}_M$,
respectively. Take the prediction process of $\mbox{\rm TX}_m$ as an
example. At the energy harvesting moment $T_{m,n}$, $\mbox{\rm
TX}_m$ harvests energy $E_{m,n}$. The length of the time slot between two consecutive energy
harvesting moments is $\mathcal{T}_{m,n}=T_{m,n}-T_{m,n-1}$. At
the present time $T$, $\mbox{\rm TX}_m$ has the amount of left
energy $E_m$. The last energy harvesting moment
before $T$ is denoted by $T_{m,0}$.

Denote the duration from $T$ until the moment the left energy is
used up as the first working time period. Then, we can obtain the
length of the first working time period $\mathbb{T}_{m,1}$ as
$\mathbb{T}_{m,1}=\frac{E_m}{P_1}$. During the first working time
period, if there is no energy arriving, the working time of
$\mbox{\rm TX}_m$ is $\mathbb{T}_{m,1}$; otherwise, the newly
harvested energy $E_{m,1}$ can be used immediately
and the length of the first working time period
changes to $\mathbb{T}_{m,2}$.

Let $ma= T+ \mathbb{T}_{m,1}-T_{m,0}$. Then, we can get the probability of no energy arriving in the first working time period as
\begin{equation}\label{eujyu4}
  PP_{1}=PP(\mathcal{T}_{m,1}\geq ma)=e^{-\lambda_{m}ma}
\end{equation}
When $\mathcal{T}_{m,1}< ma$, $E_{m,1}$ can be used during the first working time period and $\mbox{\rm TX}_m$ can work for a longer time $\mathbb{T}_{m,2}$, as $\mathbb{T}_{m,2}=\frac{E_m+E_{m,1}}{P_1}$.
Before the epoch $T+\mathbb{T}_{m,2}$, the probability of no energy arriving is
\begin{equation}\label{eqyr6}
\begin{aligned}
  PP_{2}&=PP\bigg(\mathcal{T}_{m,1}< ma, \mathcal{T}_{m,1}+\mathcal{T}_{m,2}\geq\frac{E_{m,1}}{P_1}+ma \bigg)\\
  &=PP(\mathcal{T}_{m,1}< ma)\\
  &\times PP(\mathcal{T}_{m,1}+\mathcal{T}_{m,2}-\frac{E_{m,1}}{P_1}\geq ma)\\
  &=(1-PP_1)\times \tilde{PP}_2
  \end{aligned}
\end{equation}
where $\tilde{PP}_2=PP(\mathcal{T}_{m,1}+\mathcal{T}_{m,2}-\frac{E_{m,1}}{P_1}\geq ma)$. And likewise, we can calculate the following probabilities:

 \begin{equation}\label{eqgfg}
  \mathbb{T}_{m,3}=\frac{E_m+E_{m,1}+E_{m,2}}{P_1}
\end{equation}
\begin{equation}\label{eqyr9}
\begin{aligned}
  PP_{3}&=PP\bigg(\mathcal{T}_{m,1}< ma, \mathcal{T}_{m,1}+\mathcal{T}_{m,2}<\frac{E_{m,1}}{P_1}+ma,\\
  &\mathcal{T}_{m,1}+\mathcal{T}_{m,2}+\mathcal{T}_{m,3}\geq \frac{E_{m,1}+E_{m,2}}{P_1}+ma \bigg)\\
  &=(1-PP_1-PP_2)\times \tilde{PP}_3
  \end{aligned}
\end{equation}
\qquad \qquad \qquad \qquad \qquad \qquad $\vdots$
\begin{equation}\label{eqgfg}
  \mathbb{T}_{m,n}=\frac{E_m+E_{m,1}+E_{m,2}+\cdots+E_{m,n-1}}{P_1}
\end{equation}

{\small \begin{equation}\label{eqyr11}
\begin{aligned}
  PP_{n}&=PP(\mathcal{T}_{m,1}< ma,\mathcal{T}_{m,1}+\mathcal{T}_{m,2}<\frac{E_{m,1}}{P_1}+ma,\cdots,\\
  &\mathcal{T}_{m,1}+\mathcal{T}_{m,2}+\cdots+\mathcal{T}_{m,n}\\
  &\geq\frac{E_{m,1}+E_{m,2}+\cdots+E_{m,n-1}}{P_1}+ma)\\
  &=(1-PP_1-\ldots-PP_{n-1})\\
  &\times PP(\mathcal{T}_{m,1}+\mathcal{T}_{m,2}+\cdots+\mathcal{T}_{m,n}\\
  &-\frac{E_{m,1}+E_{m,2}+\cdots+E_{m,n-1}}{P_1}>ma)\\
  &=(1-PP_1-\ldots-PP_{n-1})\times \tilde{PP}_n
  \end{aligned}
\end{equation}}
where
\begin{equation}\label{ppxn}
\begin{aligned}
  \tilde{PP}_n&=PP(\mathcal{T}_{m,1}+\mathcal{T}_{m,2}+\cdots+\mathcal{T}_{m,n}\\
  &-\frac{E_{m,1}+E_{m,2}+\cdots+E_{m,n-1}}{P_1}>ma)
    \end{aligned}
\end{equation}

Next, we will discuss the way to get
$\tilde{PP}_n$, in order to obtain the probability of no energy
arriving. Let
$X=\mathcal{T}_{m,1}+\mathcal{T}_{m,2}+\cdots+\mathcal{T}_{m,n}$,
$Y=E_{m,1}+E_{m,2}+\cdots+E_{m,n-1}$. Then, (\ref{ppxn}) turns into
 \begin{equation}\label{ppsn2}
  \tilde{PP}_n=PP \bigg(X-\frac{Y}{P_1}>ma \bigg)
\end{equation}
From \cite{rtoi}, we can get the probability density function of $X$ and $Y$ as
 \begin{equation}\label{x}
  f_X(x)=\frac{\lambda_m ^n}{(n-1)!}x^{n-1} e^{-x},\quad x\geq0
\end{equation}

\begin{equation}\label{y}
\begin{aligned}
  f_Y(y)=\frac{\sum_{k=0}^{n-1} (-1)^k \binom{n-1}{k}(C(\frac{y-(n-1)\mbox{\rm dn}_m}{\mbox{\rm up}_m-\mbox{\rm dn}_m}-k))^{n-2}}{(\mbox{\rm up}_m-\mbox{\rm dn}_m)(n-2)!},\\
  \quad (n-1)\mbox{\rm dn}_m\leq y\leq(n-1)\mbox{\rm up}_m
    \end{aligned}
\end{equation}
where
\[
C(y)=
\begin{cases}
0, &\text{if $y < 0$}\\
y, &\text{if $y \geq 0$}
\end{cases}
\]
Since $X$ and $Y$ are independent, their joint probability density
function is
\begin{equation}
f_{X,Y}(x,y)=f_X(x)f_Y(y)
\end{equation}
We can get $\tilde{PP}_n$ as
\begin{equation}\label{ppx}
\begin{aligned}
\tilde{PP}_n&=\int_{ma+\frac{(n-1)\mbox{\rm dn}_m}{P_1}}^{ma+\frac{(n-1)\mbox{\rm up}_m}{P_1}}dx\int_{(n-1)\mbox{\rm dn}_m}^{(x-ma)P_1} f_{X,Y}(x,y)dy\\
&+\int_{ma+\frac{(n-1)\mbox{\rm up}_m}{P_1}}^{+\infty}dx\int_{(n-1)\mbox{\rm dn}_m}^{(n-1)\mbox{\rm up}_m} f_{X,Y}(x,y)dy
\end{aligned}
\end{equation}
By substituting (\ref{ppx}) into (\ref{eqyr6}), (\ref{eqyr9}), and (\ref{eqyr11}), $PP_2, PP_3, \ldots, PP_{n}$ are obtained.

With $PP_1, PP_2, \ldots, PP_{n}$ at hand, we can
find out the mean of the working time variables
$\mathbb{T}_{m}=\{\mathbb{T}_{m,1},\cdots,\mathbb{T}_{m,n}\}$,
which is
{\begin{equation}
\begin{aligned}
\textsf{E}(\mathbb{T}_{m})&=\mathbb{T}_{m,1}\times
PP_1+\mathbb{T}_{m,2}\times PP_2+\\
&\cdots+\mathbb{T}_{m,n}\times PP_n
\end{aligned}
\end{equation}
$\mathbb{T}_{m,n}$ is a function of $E_{m,n}$ and $E_{m,n}$ is
random, which makes $\textsf{E}(\mathbb{T}_{m})$ a random variable.
We need to seek the mean again. We write the mean of
$\textsf{E}(\mathbb{T}_{m})$ as {\small
\begin{equation}
\begin{aligned}
\bar{\mathbb{T}_{m}}&=\textsf{E}(\textsf{E}(\mathbb{T}_{m}))=PP_1\times \mathbb{T}_{m,1}
+ PP_2 \times \bigg(\mathbb{T}_{m,1}+\frac{\textsf{E}(E_{m,1})}{P_1}\bigg)\\
&+\cdots + PP_n \times \bigg(\mathbb{T}_{m,1}+\frac{\textsf{E}(E_{m,1}+E_{m,2}+\cdots +E_{m,n})}{P_1}\bigg)\\
&=\mathbb{T}_{m,1}+\frac{PP_2+2PP_3+\cdots +(n-1)PP_n}{P_1}
\times \frac{\mbox{\rm up}_m-\mbox{\rm dn}_m}{2}\\
\end{aligned}
\end{equation}}

By substituting $P_1$ into (\ref{rate}), the rate $r_m$ is obtained. After multiplying $r_m$ with $\bar{\mathbb{T}_{m}}$, the amount of data that can be sent by $\mbox{\rm TX}_m$ is obtained (denoted by $B_m$). Note that the principle of predicting the working time does not alter with the optimal transmission power. The same prediction process can be executed in the other time slots for the ``whole transmitter''. For clarity, the above prediction process is shown in {\bf Algorithm 1}.
\begin{table}
 \centering
 {\small \begin{tabular}{l}\hline
{\bf Algorithm 1} The prediction of working time for a transmitter \\
\hline
Initialize: Set $n=0$. \\
{\bf while} $(PP_n \times  \mathbb{T}_{m,n}=0)$ or $(PP_n \times \mathbb{T}_{m,n}\geq 0.01)$ {\bf do}\\
$n=n+1;$\\
Calculate $\mathbb{T}_{m,n}, PP_n$.\\
{\bf if} $PP_n \times \mathbb{T}_{m,n}<0.01$ {\bf then}\\
Calculate $\bar{\mathbb{T}_{m}}$ for $\mbox{\rm TX}_m$.\\
{\bf end if}\\
{\bf end while}\\
\hline
\end{tabular}}
\end{table}

To do projection under unknown energy harvesting processes, the prediction of the transmission completion point is also needed. However, it is very complex and is left for future work.

\section{Simulation Results\label{sec5} }
As the performance of the proposed transmitter switching policy in terms of number of switches cannot be analyzed theoretically, a simulation example is used to illustrate it. There are four transmitters which are denoted by
$\mbox{\rm TX}_1$, $\mbox{\rm TX}_2$, $\mbox{\rm TX}_3$, and
$\mbox{\rm TX}_4$. The energy parameters are set as
$\lambda_1=1$, $\lambda_2=\frac{1}{10}$,
$\lambda_3=\frac{1}{20}$, $\lambda_4=\frac{1}{30}$, $\mbox{\rm
dn}_1=1$ mJ, $\mbox{\rm up}_1=5$ mJ, $\mbox{\rm dn}_2=20$ mJ,
$\mbox{\rm up}_2=24$ mJ, $\mbox{\rm dn}_3=100$ mJ, $\mbox{\rm up}_3=104$
mJ, $\mbox{\rm dn}_4=4$ mJ, $\mbox{\rm up}_4=44$ mJ. The amount of data to be sent to the receiver are 6000 bits. The channel parameters are
set as follows: bandwidth $B_w=1$ MHz, the path loss between the
four transmitters and the receiver are $h_1=-100$ dB, $h_2=-101$ dB,
$h_3=-102$ dB, $h_4=-103$ dB, the noise power spectral density is
$N_0=10^{-19}$ W/Hz. The transmission rates can be written as
follows:
\begin{equation}
\begin{aligned}
 r_{m}&=B_w \log_2(1+\frac{P_j h_{m}}{N_0 B_w}),\quad m=1,2,3,4\\
 r_{1}& =\log_2(1+\frac{P_j}{10^{{-3}}})\,\rm{Mbps},\\
 r_{2}& =\log_2(1+\frac{P_j}{10^{{-2.9}}})\,\rm{Mbps},\\
 r_{3}& =\log_2(1+\frac{P_j}{10^{{-2.8}}})\,\rm{Mbps},\\
 r_{4}& =\log_2(1+\frac{P_j}{10^{{-2.7}}})\,\rm{Mbps}\\
\end{aligned}
\end{equation}

According to the above simulation parameters, we can get the optimal transmission powers of the ``whole transmitter'' \cite{tsfb} as $P_1=11.2082$ mW, $P_2=11.4031$ mW, $P_3=13.7658$ mW, $P_4=16.3307$ mW, $P_5=22.8431$ mW, $P_6=36.9349$ mW, $P_7=43.9203$ mW, $P_8=50.8242$ mW. The epoches are $t_1=1588.7393$ s, $t_2=1655.3670$ s, $t_3=1846.6092$ s, $t_4=1858.2105$ s, $t_5=1859.2382$ s, $t_6=1860.5000$ s, $t_7=1861.9891$ s, $t_8=1862.0000$ s. With the optimal transmission power, until the epoch $t_8$, 2207 times of energy are harvested and utilized by the system, all the bits are sent to the receiver, and the number of switches is 82.

To show the advantage, the proposed switching policy is compared with several heuristic switching policies:

\begin{enumerate}

\item {\bf Left Energy Maximum} (EM) policy: This policy only depends on the amount of left energy $E_m$ at a switching moment. One can choose the largest $E_m$ and switch to $\mbox{\rm TX}_m$.

\item {\bf Rate Maximum} (RM) policy: As we know, for a given transmission completion time, higher rate means sending a larger amount of bits out. In the same time slot, we have $r_1>r_2>r_3>r_4$. So the switching order is $\mbox{\rm TX}_1$ first, $\mbox{\rm TX}_2$ second, $\mbox{\rm TX}_3$ third, and $\mbox{\rm TX}_4$ last.

\item {\bf Bits Maximum} (BM) policy: No matter what the switching policy is, as long as all energies are consumed, the same amount of bits is transmitted. Recall that the amount of bits sent by the $m$-th transmitter in the working time is $B_m$. We pick out the largest $B_m$ and let $\mbox{\rm TX}_m$ work.

\item {\bf Working Time Maximum} (TM) policy: For a given transmission completion time, longer working time leads to a lower number of switches. In this policy, we calculate the working time of each transmitter, pick out the longest one, and let the corresponding transmitter work.

\end{enumerate}

With the optimal transmission power, the number of
switches under the switching policies are shown in Figure \ref{a11}.
\begin{figure}[t]
\centerline{\hbox{\epsfxsize=7cm\epsfbox{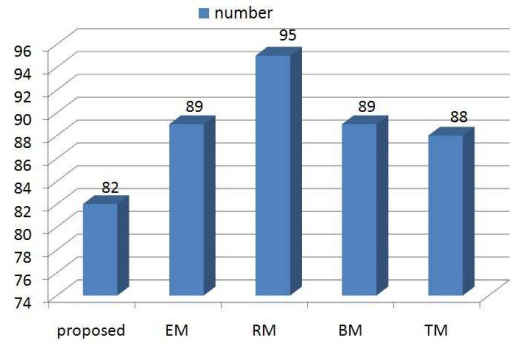}}}
\caption{Number of switches under the switching policies.} \label{a11}
\end{figure}
From this figure, we can see that the proposed switching policy leads to
a lower number of switches than all aforementioned
heuristic switching policies. It should be emphasized that the
heuristic switching policies follow the line of maximizing one of
the system parameters, which are sub-optimal.

Furthermore, we take 500 and 1000 independent runs of the same simulation and
get the average number of switches under the
switching policies respectively, as shown in Figure \ref{a22}.
\begin{figure}[t]
\centerline{\hbox{\epsfxsize=7cm\epsfbox{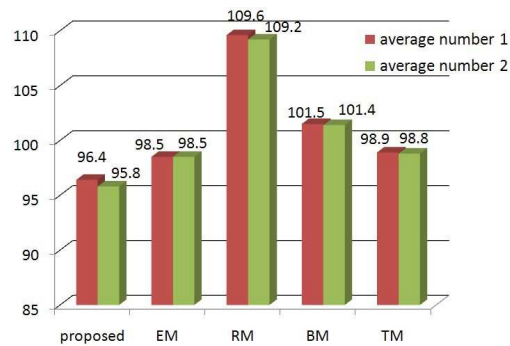}}}
\caption{Average number of switches under the switching policies.} \label{a22}
\end{figure}
This result also indicates that the average number
of switches under the proposed switching policy is the least.

\section{Conclusion\label{sec6}}
A geometric projection-based transmitter switching policy for a
communication system with multiple energy harvesting transmitters and one receiver
has been presented, which reduces the number
of switches. Transmitter switching is determined by the projection
on a line connecting the transmission start point and the
transmission completion point in the time-data plane. The proposed
switching policy leads to a lower number of
switches than several heuristic ones. In this work, we have assumed that the channel states do not vary within the transmission completion time and the channel path losses are known \cite{opsi1}. In the future, the impact of channel path losses will be incorporated in the design of transmitter switching policies.


\begin{thebibliography}{99}

\bibitem{epiw}
He S, Chen J, Jiang F, Yau DKY, Xing G, Sun Y. Energy provisioning in wireless rechargeable sensor networks. \emph{IEEE Transactions on Mobile Computing} 2013; {\bf 12}(10): 1931--1942.

\bibitem{ajas}
Alvarado U, Juanicorena A, Adin I, Sedano B, Guti\'{e}rrez I, de N\'{o} J. Energy harvesting technologies for low-power electronics. \emph{Transactions on Emerging Telecommunications Technologies} 2012; {\bf 23}(8): 728--741.

\bibitem{gw}
Guo W, Wang S. Radio-frequency energy harvesting potential: a stochastic analysis. \emph{Transactions on Emerging Telecommunications Technologies} 2013; {\bf 24}(5): 453--457.

\bibitem{opsi1}
Yang J, Ulukus S. Optimal packet scheduling in an energy harvesting communication system. \emph{IEEE Transactions on Communications} 2012; {\bf 60 }(1): 220--230.

\bibitem{ty}
Tutuncuoglu K, Yener A. Optimal transmission policies for
battery limited energy harvesting nodes. \emph{IEEE Transactions on Wireless Communications} 2012; {\bf 11}(3):1180--1189.

\bibitem{bwae}
Yang J, Ozel O, Ulukus S. Broadcasting with an energy harvesting rechargeable transmitter. \emph{IEEE Transactions on Wireless Communications} 2012; {\bf 11}(2): 571--583.

\bibitem{aue}
Antepli MA, Uysal-Biyikoglu E, Erkal H. Optimal packet scheduling on an energy harvesting broadcast link. \emph{IEEE Journal on Selected Areas in Communications} 2011; {\bf 29}(8):1721--1731.

\bibitem{eou}
Erkal H, Ozcelik FM, Uysal-Biyikoglu E. Optimal offline broadcast scheduling with an energy harvesting transmitter. \emph{EURASIP Journal on Wireless Communications and Networking} 2013.

\bibitem{obsf}
Ozel O, Yang J, Ulukus S. Optimal broadcast scheduling for an energy harvesting rechargeable transmitter with a finite capacity battery. \emph{IEEE Transactions on Wireless Communications} 2012; {\bf 11}(6): 2193--2203.

\bibitem{tweh}
Ozel O, Tutuncuoglu K, Yang J, Ulukus S, Yener A. Transmission with energy harvesting nodes in fading wireless channels: optimal policies. \emph{IEEE Journal on Selected Areas in Communications} 2011; {\bf 29}(8):1732--1743.

\bibitem{yu}
Yang J, Ulukus S. Optimal packet scheduling in a multiple access channel with energy harvesting transmitters. \emph{Journal of Communications and Networks} 2012; {\bf 14}(2):140--150.

\bibitem{otsf}
Ozel O, Yang J, Ulukus S. Optimal transmission schemes for parallel and fading Gaussian broadcast channels with an energy
harvesting rechargeable transmitter. \emph{Computer Communications} 2013; {\bf 36}(12): 1360--1372.

\bibitem{ty2}
Tutuncuoglu K, Yener A. Sum-rate optimal power policies for energy harvesting transmitters in an interference channel. \emph{Journal of Communications and Networks} 2012; {\bf 14}(2): 151--161.

\bibitem{ke}
Kashef M, Ephremides A. Optimal packet scheduling for energy harvesting sources on time varying wireless channels. \emph{Journal of Communications and Networks} 2012; {\bf 14}(2): 121--129.

\bibitem{aacu}
Gurakan B, Ozel O, Yang J, Ulukus S. Energy cooperation in energy harvesting communications. \emph{IEEE Transactions on Communications} 2013; {\bf 61}(12): 4884--4898.

\bibitem{bgd}
Blasco P, G\"{u}nd\"{u}z D, Dohler M. A learning theoretic approach to energy harvesting communication system optimization. \emph{IEEE Transactions on Wireless Communications} 2013; {\bf 12}(4): 1872--1882.

\bibitem{dg}
Devillers B, G\"{u}nd\"{u}z D. A general framework for the optimization of energy harvesting communication systems with battery imperfections. \emph{Journal of Communications and Networks} 2012; {\bf 14}(2): 130--139.

\bibitem{ouu}
Ozcelik FM, Uctu G, Uysal-Biyikoglu E. Minimization of transmission duration of data packets over an energy harvesting fading channel.
\emph{IEEE Communications Letters} 2012; {\bf 16}(12): 1968--1971.

\bibitem{xb}
Xiao C, Beaulieu NC. Node switching rates of opportunistic relaying and switch-and-examine relaying in Rician and Nakagami-$m$ fading. \emph{IEEE Transactions on Communications} 2012; {\bf 60}(2): 488--498.

\bibitem{tsfb}
Zhou F, Chen H, Zhao F. Transmission scheduling for broadcasting with two energy harvesting switching transmitters.
\emph{IET Wireless Sensor Systems} 2013; {\bf 3}(2): 138--144.

\bibitem{leht}
Lee P, Eu ZA, Han M, Tan H. Empirical modeling of a solar-powered energy harvesting
wireless sensor node for time-slotted operation. In \emph{IEEE Wireless Communications \& Networking Conference}, Cancun, Mexico, 2011; 179--184.

\bibitem{rtoi}
Gradshteyn IS, Ryzhik IM. \emph{Table of Integrals, Series, and Products (Seventh Edition)}. Academic Press: Salt Lake City, USA, 2007.

\end{thebibliography}
\end{document}